\documentclass[twocolumn,showpacs,preprintnumbers,amsmath,amssymb]{revtex4}

\usepackage{graphicx}
\usepackage{dcolumn}
\usepackage{bm}

\begin{document}

\preprint{APS/123-QED}

\title{A fluidized granular medium as an instance of the Fluctuation Theorem}

\author{Klebert Feitosa}
 \email{kfeitosa@physics.umass.edu}
\author{Narayanan Menon}%
 \email{menon@physics.umass.edu}
\affiliation{%
Department of Physics, University of Massachusetts, Amherst,
Massachusetts 01003-3720
}%

\date{April 15, 2002}

\date{\today}

\begin{abstract}
We study the statistics of the power flux into a collection of
inelastic beads maintained in a fluidized steady-state by external
mechanical driving.  The power shows large fluctuations, including
frequent large negative fluctuations, about its average value. The
relative probabilities of positive and negative fluctuations in
the power flux are in close accord with the Fluctuation Theorem of
Gallavotti and Cohen, even  at time scales shorter than those
required by the theorem.  We also compare an effective temperature
that emerges from this analysis to the kinetic granular
temperature.
\end{abstract}

\maketitle

Take a fistful of marbles in your hand and shake them vigorously.
In order to maintain the motions of the marbles, you will have to
continuously supply enough energy to them to replenish the energy
they lose when they collide inelastically with each other. This is
unlike a state of thermal equilibrium, where equal amounts of heat
flow into and out of the system of particles to the thermal bath
that sets the temperature of the system. All fluctuations in the
equilibrium state are given by the appropriate calculation within
the canonical ensemble. When this equilibrium state is perturbed
slightly (for instance, by connecting the system between two heat
baths at slightly different temperatures), the
fluctuation-dissipation theorem specifies the linear response of
the system in terms of the equilibrium fluctuations. No equivalent
framework currently exists for large departures from equilibrium.
However, a recent theorem due to Gallavotti and Cohen
{\cite{GC95JStatPhys,GC95PRL} takes an important step in this
direction. Inspired by an observation made in a simulation of a
sheared hard-sphere fluid \cite{CohenEvansMorriss}, they proved a
very general result regarding the entropy flux into a system
maintained in a nonequilibrium steady state by a time-reversible
thermostat. If dynamics in the system are chaotic \cite{CH}, then
\begin{eqnarray}
\Pi(\sigma_{\tau})/\Pi(-\sigma_{\tau})=exp(\sigma_{\tau}\tau),
\label{equ:1}
\end{eqnarray}
where $\Pi(\sigma_{\tau})$ is the probability of a fluctuation of
amplitude $\sigma_{\tau}$ in the rate of entropy production,
computed over a time $\tau$ that is much longer than any
microscopic time-scale of the system.

The Fluctuation Theorem (FT) embodied in eqn. (\ref{equ:1}) above,
is a remarkably strong statement, with no adjustable
system-dependent parameters. The significance and realm of
validity of this result are being explored theoretically in
various directions.  The theorem has been shown to be true under
Langevin dynamics \cite{Kurchan,Leibowitz&Spohn}, thus broadening
greatly the range of applicability of systems from only
Hamiltonian systems. The theorem has been shown to be equivalent
to the fluctuation-dissipation theorem in the limit of small
forces \cite{Gallavotti96}.   The FT has also been shown
\cite{Rey-Bellet} for a chain of anharmonic oscillators to produce
the Green-Kubo formula for thermal conductivity.  Recent work
\cite{Crooks99} also shows the connection between the FT and a new
thermodynamic identity \cite{Jarzynski97} that relates equilibrium
free energy differences to nonequilibrium work.  Thus, while eqn.
(\ref{equ:1}) may at first sight appear to be a rather formal
mathematical relationship, it has a very nontrivial physical
significance.

Explicit realizations of the FT have been demonstrated in
simulations of sheared fluids \cite{CohenEvansMorriss}, electrical
conductivity in an array of fixed scatterers \cite{Bonetto},
shell-models of turbulence \cite{BiferalePierotti98}, a model of a
granular medium and a Burridge-Knopoff type block-spring model
\cite{Aumaitre}. Clear experimental demonstrations of the FT have
not previously been achieved, but the two previous attempts have
been very instructive in the difficulties involved.  Goldburg et
al. \cite{Goldburg} studied the fluctuations in electrical current
required to maintain in steady-state a liquid-crystal film driven
into chaotic convectional motion by a constant transverse voltage.
No direct test of equation (\ref{equ:1}) was possible due to the
fact that negative fluctuations in power are exceedingly rare in a
macroscopic system.  Ciliberto and Laroche \cite{Ciliberto}
studied temperature fluctuations in a Rayleigh-Benard cell arguing
that the temperature variations at a point in a chaotic flow are
representative of the heat fluxes to which the FT applies; they
side-step the difficulty of measuring negative power fluctuations
encountered in \cite{Goldburg} by using as the variable $\sigma$
the local temperature minus the mean temperature in the cell. They
find that the ratio on the left side of eqn. (\ref{equ:1}) is
indeed exponential in the amplitude of the fluctuations, but their
results are less conclusive on the $\tau$ dependence.

Informed by these previous studies, we are led to consider in this
article a granular medium made up of macroscopic spherical beads
in which we are able to directly measure energy fluxes. Since
granular systems typically contain much fewer particles than
molecular systems, and since relevant experimental situations are
often strongly driven non-equilibrium states, these fluxes
naturally undergo large fluctuations, including the negative
fluctuations that play an important role in the FT. These
attributes of our experimental system allow us to make a very
substantive test of the result in equation (\ref{equ:1}).
\begin{figure}
\includegraphics{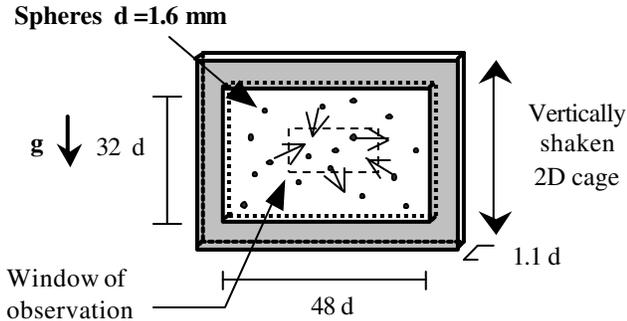}
\caption{\label{cartoon} Sketch of the experimental cell. The
dashed rectangle is a window measuring $10d\text{x}21d$, fixed in
the laboratory frame, in which we study the flux of kinetic
energy.}
\end{figure}

We make measurements of power fluctuations in the geometry
sketched in Figure \ref{cartoon}. Spherical glass beads (diameter
$d=1.6~\text{mm}$; mass $m=5.24~mg$) are held in a rectangular
cage in the vertical plane and agitated strongly by vertical
vibrations. For the measurements we report here, the frequency and
amplitude of vibration are held fixed at $60~\text{Hz}$ and $2.6
d$. We take video frames every $0.5~\text{ms}$ from which the
instantaneous positions of the particles are determined to a
precision of $0.025 d$, and their velocities are thence inferred.
We measure the time-dependence of the total energy and fluxes into
a subsystem defined by the window indicated by dashed lines in
Fig. \ref{cartoon}.  We quantify the power $P(t)$ sustaining the
dynamics in this subsystem by finding the sum of two measured
quantities. The first is the work done on the particles by
gravity, which we obtain from the difference in potential energy
of the particles in the window from one frame to the next; this
has a zero time-average. The second, and much larger, term is the
flux in kinetic energy through the dashed lines in Fig.
\ref{cartoon} due to particles entering or exiting the window. We
neglect the contributions to the $P(t)$ from the flux of
rotational kinetic energy. We note that these data are taken in a
dilute regime (area fraction $\sim 13.8\%$), so that $P(t)$ is
well-approximated by measuring only the contribution from the
streaming term, i.e. ignoring the rare contributions to the energy
flux from particles at the edges of the subsystem that suffer
collisions with particles outside the subsystem.

In the subsystem we consider, the fluctuation over a time $\tau$
of the total mechanical energy is $\Delta E=D+P$, where $D$ is the
dissipation due to inelastic collisions between particles, and $P$
is the power driving the subsystem, as described above. In steady
state, the time-average $\overline{\Delta E}$ is zero, so that
$P(t)$ necessarily has a positive time-average value,
$\overline{P}$, which must exactly balance the time-average of the
dissipation $\overline{D}$. In theoretical discussions of the FT,
the entropy production rate, $\sigma$, is identified with the
phase space contraction rate. Since this is not an experimentally
accessible quantity, we have chosen to make a correspondence
between $\sigma(t)$ and $P(t)/T_{eff}$, where $T_{eff}$ is the
effective temperature of the system. The choice of what
temperature to ascribe to the system is not obvious; we can avoid
this issue in constructing the left side of equation
(\ref{equ:1}), since it is a ratio between the probabilities
positive and negative values of the same variable. However, we
must confront this issue in discussing the right side, since an
energy scale will be required to make non-dimensional the quantity
in the exponent. We defer this discussion to later in this
article.

In Figure \ref{Pwr_time}, we display a sample of the
time-variation of the normalized power, $p(t)=P/\overline{P}$ to
indicate that this quantity fluctuates strongly about its mean
($\bar{p}=1$), including making several negative excursions. We
emphasize that the occurrence of frequent negative fluctuations is
not because we are operating at small driving forces, close to
thermal equilibrium; on the contrary, the mean kinetic energy is
on the order of $10^{16}~k_{B}T$, where $T$ is the ambient
temperature that controls the microscopic degrees of freedom
internal to the particles.

With $p(t)$ being identified as the variable of interest, the
version of the Fluctuation relation that we test here is
\begin{eqnarray}
\Pi(p_{\tau})/\Pi(-p_{\tau})=exp(p_{\tau}\tau\overline{P}/T_{eff}).
\label{equ:2}
\end{eqnarray}

Before embarking on the data analysis required to test equation
(\ref{equ:2}), we note a few points that are relevant to the
implications of any results that emerge from such an analysis:

\begin{itemize}
\item The dynamics in the experimental system are not
time-reversible unlike \cite{GC95JStatPhys,GC95PRL} nor do they
satisfy microscopic detailed balance unlike the systems of
\cite{Kurchan,Leibowitz&Spohn}.


\begin{figure}
\includegraphics{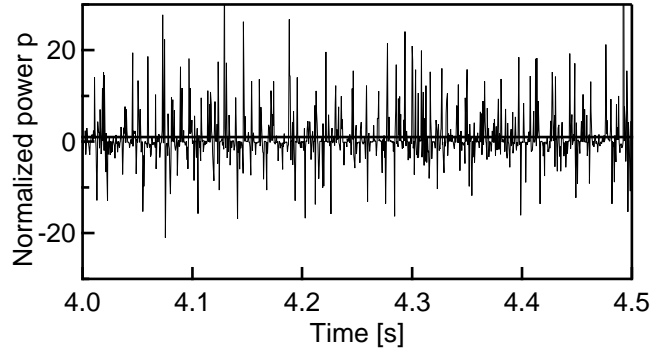}
\caption{\label{Pwr_time} A sample ($\sim0.6\%$) of the normalized
power trace for the subsystem in consideration. The horizontal
line shows the average power, $\bar{p}=1$.}
\end{figure}


\item We study an open subsystem of the entire nonequilibrium
system rather than the global power injection.  The FT is proven
for global fluxes and local versions of equation (\ref{equ:1})
have not been proved except for special cases
\cite{GallavottiLocal}.

\item To prefigure the data analysis to follow, we find that we
are not able to satisfy the condition that $\tau$ is much greater
than the dynamical time-scales in the system. However, we find
that the statistics of $p(t)$ are strongly non-gaussian for
smaller $\tau$ and therefore provide a very demanding test of
equation (\ref{equ:2}).

\end{itemize}

In order to construct the left side of equation (\ref{equ:2}), we
first partition the time trace of $p(t)$ into non-overlapping bins
of duration $\tau$ and compute the average power in these bins,
$p_{\tau}(t)=\frac{1}{\tau}\int_{t}^{t+\tau}p(t')dt'$. The
probability distribution $\Pi(p_{\tau})$ is displayed in figure
\ref{hist_p} for $\tau =0.5, 1, 2, 4, 8~\text{and}~16~\text{ms}$
for a data set constructed from a sequence of 170,000 video
frames. The data show large, non-gaussian deviations from the mean
value of
\begin{figure}
\includegraphics{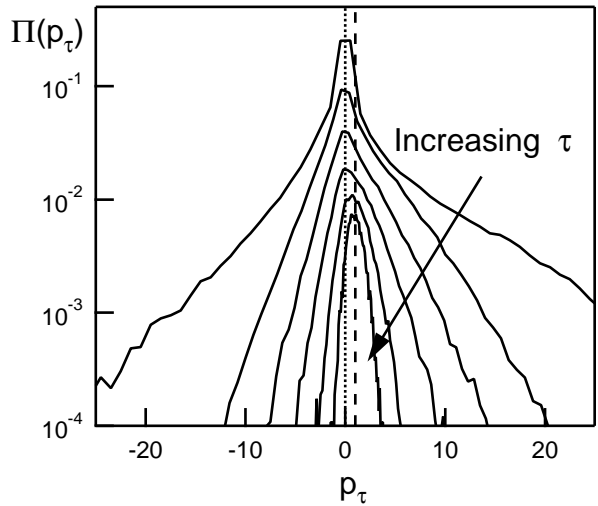}
\caption{\label{hist_p} Probability distribution, $\Pi(p_{\tau})$
of the binned power in time bins $\tau=0.5, 1, 2, 4, 8~\text{and}~
16~\text{ms}$. The distributions are displaced vertically for
clarity. The average power for all distributions is
$\bar{p}_{\tau}=1$, indicated by the dashed vertical line.}
\end{figure}
$\bar{p}_{\tau}=1$, with substantial negative tails. The shapes of
the distributions vary with $\tau$, thus each distribution
provides an independent test of eqn. (\ref{equ:2}). As noted
earlier, even with the large data set that we consider, the values
of $\tau$ we are able to access are not truly long compared to the
dynamical time-scales of the system, as required by the FT. For
comparison, the mean free time is $4~\text{ms}$ and the time for a
particle to diffuse across the window is $10~\text{ms}$.  The
autocorrelation function of $p(t)$ shows weak oscillations with a
period of $8~\text{ms}$, which corresponds to the second harmonic
of the drive frequency: this is due to energy injection from the
top and bottom walls in opposite phases of the drive cycle.
\begin{figure}
\includegraphics{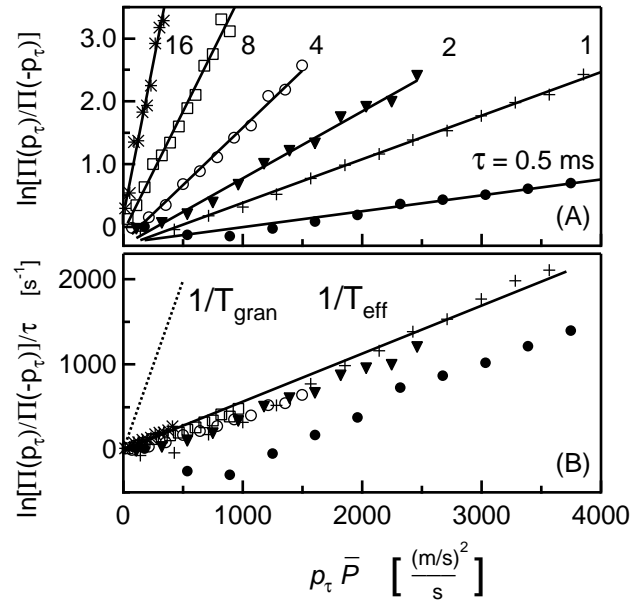}
\caption{\label{lnP} (A) $ln[\Pi(p_{\tau})/\Pi(-p_{\tau})]$ versus
$p_{\tau}\overline{P}$ for $\tau$ ranging from 0.5 to 16 ms. (B)
$ln[\Pi(p_{\tau})/\Pi(-p_{\tau})]/\tau$ versus $p_{\tau}\bar{P}$
($\overline{P}=356~m^2s^{-3}$). The solid line shows the slope of
the collapsed curves. A dashed line of slope $1/T_{gran}$ is drawn
for comparison.}
\end{figure}

In Fig. \ref{lnP}A we show for the probability distributions of
Fig. \ref{hist_p}, the variation of the ratio
$ln[\Pi(p_{\tau})/\Pi(-p_{\tau})]$ with $p_{\tau}$.  This ratio is
a straight line, as predicted from the exponential dependence in
eqn. (\ref{equ:1}).  The only deviations from linearity are the
points close to $p=0$ on the $\tau = 0.5$ and $1~\text{ms}$ data
sets. To study the $\tau$-dependence implied by the FT, in Fig.
\ref{lnP}B we plot the dependence on $\tau$ of the quantity
$ln[\Pi(p_{\tau})/\Pi(-p_{\tau})]/\tau$.  For all values of
$\tau$, we obtain the same slope (similar results were observed in
the simulations in Ref \cite{Aumaitre}), and for large $\tau$, all
the data sets collapse on a line passing through the origin, as
predicted by eqn. (\ref{equ:2}).  This data collapse indicates a
significant degree of agreement with the FT since each of the
lines derive from differently shaped $\Pi(p_{\tau})$. Furthermore
the linear dependence of the ordinate in Figure 4 extends to
values of $p_{\tau}$ well beyond the mean so that it seems
unlikely that the linearity is merely the leading order behavior
in a more complicated functional dependence on $p_{\tau}$.

Finally, we consider the slope of the lines in Fig. \ref{lnP}; as
discussed earlier, had we used the rate of entropy flow into the
system these lines are predicted to have a universal slope of
unity.  With our choice of variable, viz. the power flux, the
lines in Fig. \ref{lnP} have a slope with the dimension of an
inverse energy/mass, which should be given by the inverse of the
effective temperature, $T_{eff}$, of the system. $T_{eff}$ bears
no relation to the ambient temperature.  A plausible candidate for
a temperature scale, however, is the so-called ``granular
temperature'' $T_{gran}= 1/2\langle v^{2}\rangle$, a purely
kinetically defined temperature that we have previously studied
\cite{Rouyer2000,Feitosa2002} in the same experimental geometry.
The dotted line in Fig. \ref{lnP} has a slope of $1/T_{gran}$:
$T_{eff}$ and $T_{gran}$ are clearly unequal. To study how they
are related to one other, in Figure \ref{Inv_slope} we display the
variation of $T_{eff}$ and $T_{gran}$ as a function of the number
of particles in the cell, keeping all other parameters of the
driving mechanism fixed. As the number of particles increases,
$T_{gran}$ decreases due to an increased rate of inelastic
collisions. (We make use of the fact that the slopes of the lines
in Fig. \ref{lnP} do not depend on $\tau$, and plot only the
values of the slope for $\tau=1~\text{ms}$, for which more modest
statistics are sufficient).  As can be seen, $T_{eff}$ and
$T_{gran}$ differ in magnitude by about a factor of 8, however
they both decrease similarly as the number of particles is varied.
\begin{figure}
\includegraphics{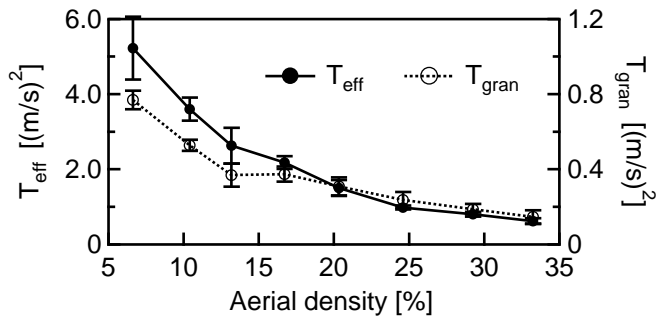}
\caption{\label{Inv_slope} $T_{eff}$ (left axis), defined as the
inverse of the slope in a graph such as Figure 4B, and $T_{gran}$
(right axis), the granular temperature ($1/2\langle
v^{2}\rangle$), as a function of area fraction. $T_{eff}$ is
numerically larger but follows a similar trend.}
\end{figure}

Prior to the proof of the FT presented in
{\cite{GC95JStatPhys,GC95PRL}, an even stronger statement was
proven by Evans and Searles {\cite{EvansSearlesPRA94} in which
they showed that eqn. (\ref{equ:1}) was true at all $\tau$. This
statement, however, applies only to systems that start in
equilibrium and are perturbed by a time-independent force. A
recent experiment {\cite{Wang02} on a colloidal particle pulled
through a fluid by an optical trap tests this transient
fluctuation theorem by examining an integral version of eqn.
(\ref{equ:1}). The transient fluctuation theorem of Evans and
Searles, however, is not expected to apply to the nonequilibrium
steady states that we probe here \cite{NarayanDhar03,
vanZonCohen2003}. The fact that eqn. (\ref{equ:2}) appears to hold
at small $\tau$ in our experimental situation should be viewed as
a characteristic of the structure of nonequilibrium stationary
states that is specific to fluidized granular media.

As discussed earlier, the experimental system is not an idealized
instance of the FT: the conditions of time-reversibility or
detailed balance do not obtain in the experiment, the variable
$p(t)$ is not an entropy generation rate, we only study a small,
open, subsystem rather than the entire system, and we are not able
to go to extremely long timescales. Despite this, we find
excellent agreement between our results and the predictions of the
FT. This gives hope that these ideas can be extended to some
situations where the conditions of proof are not met.  The
interpretation of the effective temperature in this analysis
remains an open question {\cite{GallavottiCohen03}. We speculate
that $T_{eff}$ is potentially valuable in regimes where the
granular temperature is no longer expected to be useful.

It is a pleasure to thank O. Narayan, J.S. Urbach, L. Rey-Bellet,
and most especially, Giovanni Gallavotti for educative
conversations. We also thank O.N. and G.G. for their comments on
our manuscript. NM is also grateful to R. Kotecky and other
participants at the MPI Workshop on Mathematical Aspects of
Material Science who suggested we explore connections of our work
with the FT. We received financial support from NSF DMR CAREER
9878433.



\begin{thebibliography}{14}
\bibitem{CH}
More precisely, it must satisfy the Chaotic Hypothesis, which is a
defining criterion of an Anosov system, see [1].
\end{thebibliography}

\end{document}